\documentclass[journal]{IEEEtran}

\usepackage{setspace}
\usepackage{epsfig}  
\usepackage{graphicx}  
\usepackage{amsmath}  
\usepackage{amssymb}  
\usepackage{lscape}  
\usepackage[justification=raggedright,font=small]{caption}	
\usepackage[labelformat=simple]{subcaption}

\usepackage{color}
\usepackage{cite}
\usepackage{mathrsfs}
\usepackage{mathtools}
\usepackage{amsthm}
\usepackage{array}
\usepackage{booktabs}
\usepackage{cases}
\usepackage{bm}
\usepackage{textcomp}
\newcommand{\textapprox}{\raisebox{0.5ex}{\texttildelow}}

\usepackage[framemethod=tikz,xcolor=true]{mdframed}

\newmdenv[%
leftmargin=0in,
rightmargin=0in,
tikzsetting={draw=black, line width=2.0pt}%
]{emphasize}%

\ifCLASSINFOpdf
\else
\fi

\begin{document}

\title{An Introduction to the Transmon Qubit for Electromagnetic Engineers}

\author{Thomas E. Roth~\IEEEmembership{Member,~IEEE}, Ruichao Ma, and Weng C. Chew~\IEEEmembership{Life Fellow,~IEEE}
\thanks{This work was supported by NSF ECCS 169195, a startup fund at Purdue University, and the Distinguished Professorship Grant at Purdue University.
	
Thomas E. Roth is with the School of Electrical and Computer Engineering, Purdue University, West Lafayette, IN 47907 USA. Ruichao Ma is with the Department of Physics and Astronomy, Purdue University, West Lafayette, IN 47907 USA. Weng C. Chew is with the Department of Electrical and Computer Engineering, University of Illinois at Urbana-Champaign, Urbana, IL 61801 USA and the School of Electrical and Computer Engineering, Purdue University, West Lafayette, IN 47907 USA (contact e-mail: rothte@purdue.edu).

This work has been submitted to the IEEE for possible publication. Copyright may be transferred without notice, after which this version may no longer be accessible.}
}


\maketitle

\begin{abstract}
One of the most popular approaches being pursued to achieve a quantum advantage with practical hardware are superconducting circuit devices. Although significant progress has been made over the previous two decades, substantial engineering efforts are required to scale these devices so they can be used to solve many problems of interest. Unfortunately, much of this exciting field is described using technical jargon and concepts from physics that are unfamiliar to a classically trained electromagnetic engineer. As a result, this work is often difficult for engineers to become engaged in. We hope to lower the barrier to this field by providing an accessible review of one of the most prevalently used quantum bits (qubits) in superconducting circuit systems, the transmon qubit. Most of the physics of these systems can be understood intuitively with only some background in quantum mechanics. As a result, we avoid invoking quantum mechanical concepts except where it is necessary to ease the transition between details in this work and those that would be encountered in the literature. We believe this leads to a gentler introduction to this fascinating field, and hope that more researchers from the classical electromagnetic community become engaged in this area in the future.
\end{abstract}


\IEEEpeerreviewmaketitle

\section{Introduction}
\label{sec:intro}
\IEEEPARstart{Q}{uantum} computing and quantum communication are becoming increasingly active research areas, with recent advances in the field breaking into a ``quantum advantage'' where it has been claimed that quantum hardware has outperformed what is possible with the most powerful state-of-the-art classical hardware \cite{arute2019quantum,zhong2020quantum}. Although many hardware platforms are still being pursued to implement quantum information processing, superconducting circuit architectures are one of the leading candidates \cite{blais2004cavity,blais2007quantum,gu2017microwave}. These systems, also often called \textit{circuit quantum electrodynamics (QED) devices}, utilize the quantum dynamics of electromagnetic fields in superconducting circuits to generate and process quantum information.

Circuit QED devices are most often formed by embedding a special kind of superconducting device, known as a Josephson junction, into complex systems of planar microwave circuitry (also made from superconductors) \cite{blais2004cavity}. At a high-level, a Josephson junction can be viewed as being synonymous to a nonlinear inductor. By engineering Josephson junctions into various circuit topologies, the nonlinear spectra of the resulting circuit can be optimized to form the basis of a qubit with individually addressable quantum states. Using superconductors for the surrounding microwave circuitry also leads to close to zero dissipation. This allows the typically fragile quantum states involved to survive for long enough times that planar ``on-chip'' realizations of the fundamental interactions of light and matter necessary to create and process quantum information can be leveraged at microwave frequencies \cite{blais2004cavity,blais2007quantum,gu2017microwave}. Using these tools and the flexibility of these architectures, circuit QED designs have been implemented for a wide range of quantum technologies including analog quantum computers \cite{ma2019dissipatively}, digital or gate-based quantum computers \cite{arute2019quantum,kandala2017hardware,barends2014superconducting}, single photon sources \cite{houck2007generating,zhou2020tunable,lang2013correlations}, quantum memories \cite{reagor2016quantum,sardashti2020voltage}, and components of quantum communication systems \cite{leung2019deterministic}.

Circuit QED systems have garnered such a high degree of interest in large part because of the engineering control in the design and operation of these systems, as well as the strength of light-matter coupling that can be achieved. The engineering control is due to the many types of qubits that can be designed to have desirable features using different configurations of Josephson junctions \cite{gu2017microwave,vion2002manipulating,manucharyan2009fluxonium,koch2007charge}. Further, the operating characteristics of these qubits can be tuned \textit{in situ} by electrically biasing them, allowing for dynamic reconfiguration of the qubit that is not possible with fixed systems like natural atoms and ions \cite{gu2017microwave}. Additionally, the macroscopic scale of circuit QED qubits and the high confinement of electric and magnetic fields possible in planar transmission lines can be leveraged to achieve exceptionally strong coupling \cite{devoret2007circuit}. As a result, circuit QED systems have achieved some of the highest levels of light-matter coupling strengths seen in any physical system to date, providing an avenue to explore and harness untapped areas of physics \cite{kockum2019ultrastrong}. Finally, many aspects of the fabrication of these systems can follow from established semiconductor fabrication techniques \cite{gu2017microwave}. However, there remain many open challenges and opportunities related to fabrication of these devices to continue to improve their performance \cite{krantz2019quantum}.

\begin{figure}[t]
	\centering
	\includegraphics[width=0.7\linewidth]{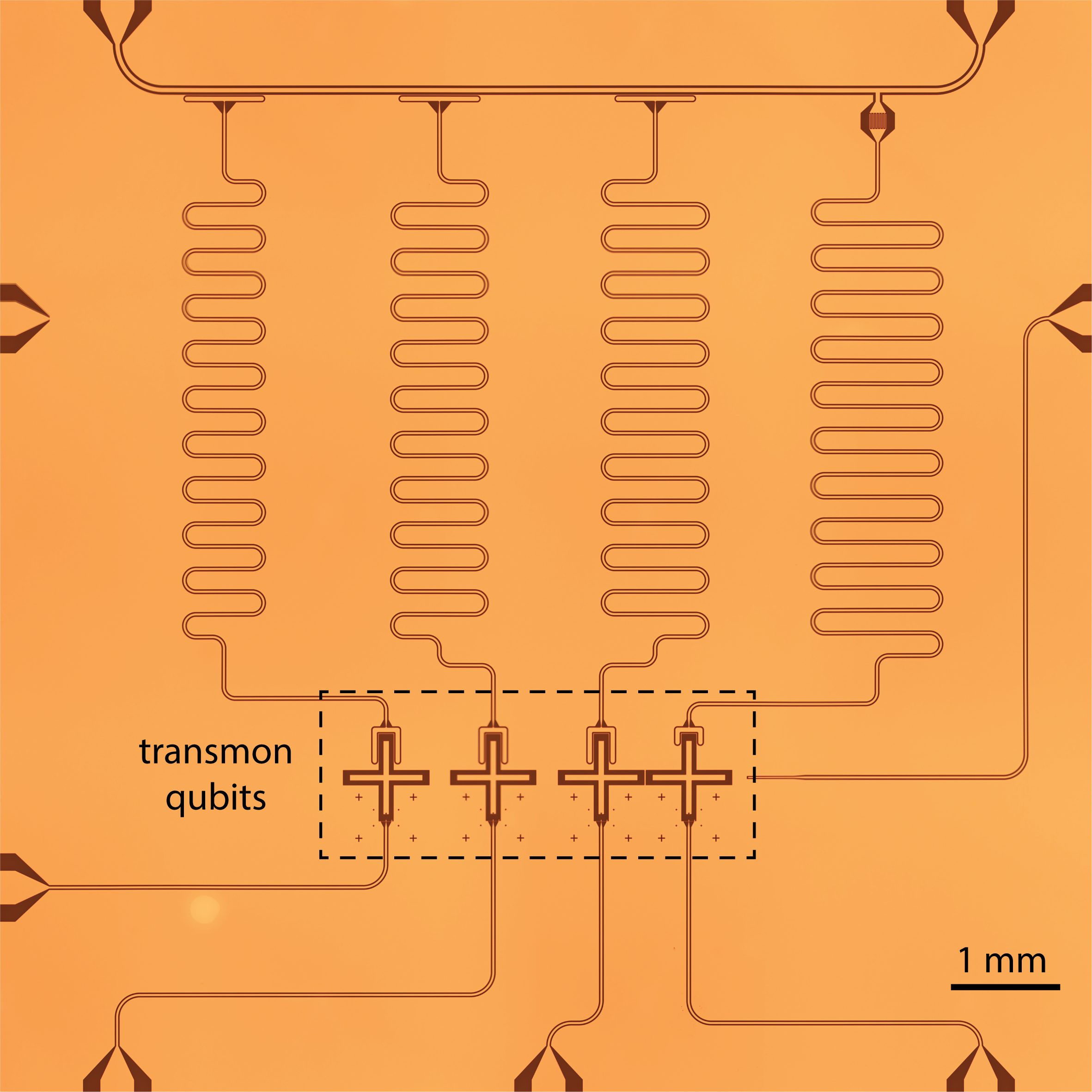}
	\caption{Optical image of a typical circuit QED device with four transmon qubits and the accompanying on-chip microwave network needed to operate the device. Each arm of the transmons are \textapprox300 $\mu$m long and form a \textapprox70 fF capacitance to ground. The main operating frequencies of the transmons in this device are around 5 GHz, but this can be tuned \textit{in situ} between approximately 4 to 6 GHz. The surrounding microwave transmission lines are used to control the transmons and monitor their quantum states through microwave spectroscopy.}
	\label{fig:full-device}
\end{figure}

Of the many superconducting qubits that have been designed to date, the \textit{transmon} has become one of the most widely used since its creation and forms a key part of many scalable quantum information processing architectures using superconducting circuits \cite{koch2007charge,arute2019quantum}. An example of the typical on-chip features needed to operate a circuit QED device with four transmon qubits is shown in Fig. \ref{fig:full-device}. The features of this device will be discussed in more detail throughout this review.

In an effort to help the classical electromagnetic community become engaged in circuit QED research, we present a review of the essential properties of the transmon qubit. Our review is presented in a manner that should be largely accessible to individuals with a limited background in quantum mechanics. However, to achieve a deeper understanding of the quantum concepts discussed in this review, readers can also consult \cite{chew2021qme-made-simple}. We begin this review by introducing the basic aspects of superconductivity that are needed to understand the operation of a Josephson junction, before building an intuitive picture for understanding the dynamics of these systems. Following this, we present the development of the transmon qubit in a historical context by considering how it evolved from earlier qubits. We then review a common model used to describe the coupling of a transmon to other microwave circuitry. This model is a typical starting point for analyzing circuit QED systems, and appears frequently in the literature. Finally, we conclude by mentioning a few areas where electromagnetic engineers can help advance this exciting new field.

\section{Superconductivity for Circuit QED}
The phenomenon of superconductivity plays a central role in the operation of all circuit QED systems. Fortunately, the physical effects of superconductivity that are needed in understanding circuit QED systems does not require a full microscopic theory to be used. As a result, only a minimal knowledge of superconductivity is needed to understand the general physics of these systems. We review only the essential properties of superconductivity in this section. A more detailed introduction to superconductivity in the context of circuit QED systems can be found in \cite{vool2017introduction,langford2013circuit,girvin2011circuit}. More complete details can be found in common textbooks; e.g., \cite{tinkham2004introduction}.

Typically, quantum effects are only observable at microscopic levels due to the fragility of individual quantum states. To observe quantum behavior at a macroscopic level (e.g., on the size of circuit components), a strong degree of coherence between the individual microscopic quantum systems must be achieved \cite{girvin2011circuit}. One avenue for this to occur is in superconductors cooled to extremely low temperatures \cite{kockum2019quantum}. At these temperatures, electrons in the material tend to become bound to each other as \textit{Cooper pairs}, which become the charge carriers of the superconducting system \cite{vool2017introduction,girvin2011circuit,langford2013circuit}. These Cooper pairs are quasiparticles formed between two electrons with equal and opposite momenta, including the spins of the electrons \cite{langford2013circuit}. Many of the properties of superconductivity are able to occur due to the pairing of the electrons.

One particularly important result is that the Pauli exclusion principle does not apply to Cooper pairs, allowing them all to ``condense'' into a single quantum ground state (often referred to as a condensate) \cite{langford2013circuit}. In addition to this, Cooper pairs are highly resilient to disturbances caused by scattering events (e.g., scattering from atomic lattices), making the overall quantum state describing the Cooper pair very stable. It is this remarkable stability that allows the quantum aspects of a superconductor to be characterized by a small number of \textit{macroscopic quantum states} \cite{langford2013circuit}. As a result, the detailed tracking of the states of individual Cooper pairs is unnecessary and a collective wavefunction can be used instead. Importantly, these collective wavefunctions have the necessary degree of coherence over large length scales to make observing macroscopic quantum behavior possible. Circuit QED systems interact with these macroscopic quantum states using microwave photons and other circuitry \cite{gu2017microwave}.

To ensure that circuit QED systems can interact with the desired macroscopic quantum states, the superconducting system must be cooled to an appropriately low temperature. To observe basic superconducting effects, the relevant materials typically need to be cooled to temperatures on the order of several kelvin \cite{girvin2011circuit}. However, circuit QED systems must operate at even lower temperatures than this to ensure that the thermal noise in the setup does not ``swamp'' the very low energy microwave photons used in these systems. As a result, circuit QED devices often must operate at temperatures on the order of 10 mK, which is achievable with modern dilution refrigerators \cite{vool2017introduction}. Although achievable, this impacts the cost of the overall system and the design constraints of auxiliary components such as microwave sources, circulators, mixers, and amplifiers, to name a few.

\section{Physics of a Josephson Junction}
\label{sec:jj-physics}
With the basic properties of superconductors understood, we can begin to consider more complicated superconducting systems that are useful in designing qubits. Of utmost interest is a \textit{Josephson junction}, which is typically formed by a thin insulative gap (on the order of a nm thick) between two superconductors \cite{kockum2019quantum}. Although many different materials can be used to form a Josephson junction, the most widely used in circuit QED systems is Al/AlOx/Al. As we will see, this superconductor-insulator-superconductor ``sandwich'' exhibits a nonlinear I-V relationship, which is a key property needed in designing most qubits.

This nonlinearity arises because the insulative gap of the Josephson junction acts as a barrier to the flow of Cooper pairs. However, as is commonly found at potential barriers, the wavefunctions of the Cooper pairs can extend into the insulative gap. By keeping the thickness of the gap small enough, the wavefunctions between the two superconductors can overlap, allowing for interactions between the two superconducting regions. Due to this interaction, it is possible for Cooper pairs to coherently tunnel between the two superconducting regions without requiring an applied voltage \cite{tafuri2019introductory}. The resulting tunneling supercurrent can be shown to have a nonlinear dependence on the voltage over the junction that is synonymous to the behavior of a nonlinear inductor. 

\subsection{Hamiltonian Description of a Josephson Junction}
\label{subsec:jj-hamiltonian}
Since we will be considering quantum aspects of a Josephson junction later, it is desirable to consider a Hamiltonian mechanics description of the Josephson junction (for basic introductions to Hamiltonian mechanics in quantum theory, see \cite{chew2016quantum,chew2016quantum2,chew2021qme-made-simple}). This amounts to expressing the total energy of the junction in terms of \textit{conjugate variables}. These conjugate variables vary with respect to each other in a manner to ensure that the total energy of the system is conserved \cite{chew2021qme-made-simple}. For the Josephson junction system, the conjugate variables are the Cooper pair density difference $n$ and the Josephson phase $\varphi$. It is important to emphasize that voltage and current are not suitable conjugate variables, which is why these variables are not typically found in the quantum treatments of Josephson junctions. Initially considering the classical case, $n$ and $\varphi$ are real-valued deterministic numbers. 

We will now assume that we have two isolated, finite-sized superconductors that are connected to each other by a Josephson junction. For the classical case, $n$ is the net density of Cooper pairs that have tunneled through the Josephson junction relative to some equilibrium level \cite{vool2017introduction,girvin2011circuit,langford2013circuit}. The Josephson phase $\varphi$ of the junction corresponds to the phase difference between the macroscopic condensate wavefunctions of the two superconductors \cite{vool2017introduction,girvin2011circuit,langford2013circuit}.

The Hamiltonian of the Josephson junction can be found by considering the total energy of the junction in terms of an effective inductance and capacitance expressed in terms of $n$ and $\varphi$. For most Josephson junctions, the effective inductance is the dominant effect. The form of the inductance can be readily inferred from the two Josephson relations, given as
\begin{numcases}
\displaystyle I=I_c \sin{\varphi},  \label{eq:jr1} 
\\
\displaystyle \frac{\partial \varphi}{\partial t} = \frac{2e}{\hbar} V, \label{eq:jr2}
\end{numcases}
where $I$ is the supercurrent flowing through the junction and $V$ is the voltage across the junction \cite{girvin2011circuit}. Further, $I_c =  2e E_J /\hbar$ is the critical current of the junction that characterizes the maximum amount of current that can coherently tunnel through the junction (i.e., exhibiting no dissipation). Finally, $E_J$ is the Josephson energy, which measures the energy associated with a Cooper pair tunneling through the junction. In terms of the Josephson inductance $L_J$ (the minimum inductance of the junction), $E_J = L_J I_c^2$ follows a typical circuit theory form.

The relations given in (\ref{eq:jr1}) and (\ref{eq:jr2}) can be used to derive the form of the nonlinear Josephson inductance by noting that
\begin{align}
\frac{\partial I}{\partial t} = \frac{\partial I}{\partial \varphi} \frac{\partial \varphi}{\partial t}  = I_c \frac{2e }{\hbar}   \cos(\varphi) \, V. 
\end{align}
Rearranging this in the form of an I-V relation for an inductor shows that the tunneling physics results in an effective inductance given by $L(\varphi) = L_J/\cos{\varphi}$, where $L_J = \hbar/(2eI_c)$ \cite{girvin2011circuit}. Overall, the energy associated with the inductance needed in the Hamiltonian for the Josephson junction is typically given as $H_L = -E_J\cos{\varphi}$ \cite{girvin2011circuit}.

Due to the physical arrangement of a Josephson junction, there is also a stray ``parallel plate'' capacitance that impacts the total energy of the junction. The energy is found by first noting that the total charge $Q$ ``stored'' in the junction capacitance is $2en$, where $2e$ is the charge of a single Cooper pair. Considering the charging energy of a single electron is $E_C = e^2/2C$, the total capacitive energy of the junction capacitance, $Q^2/2C$, can be written as $H_C = 4 E_C n^2$ \cite{girvin2011circuit}. Depending on how the Josephson junction will be used in a circuit, this small stray capacitance may be negligible so that the junction can be considered to be purely a nonlinear inductor. We include the stray capacitance here since a similar term will be important in considering transmon qubits later.

Combining the results for the dominant inductive and stray capacitive energy, the Hamiltonian for the Josephson junction system is given by
\begin{align}
H = H_C + H_L = 4E_C n^2 - E_J \cos\varphi.
\label{eq:classical-JJ-Hamiltonian}
\end{align}
From a circuit theory viewpoint, this Hamiltonian describes the total energy of a parallel combination of a linear capacitor and a nonlinear inductor. Since we have expressed the total energy of our system in terms of conjugate variables, we can use the principle of conservation of energy encoded in Hamilton's equations to find the equations of motion for this system (see \cite{chew2021qme-made-simple} for an example of this process). More importantly, finding this Hamiltonian plays a key role in developing a quantum theory for this system. 

Although the circuit theory viewpoint of (\ref{eq:classical-JJ-Hamiltonian}) can be useful, the dynamical variables $n$ and $\varphi$ are not very intuitive for solving circuit problems. Fortunately, there is an intuitive mechanical system that obeys the same dynamical equations as a Josephson junction, making it a useful analogy for understanding Josephson junction dynamics (see ``Mechanical Equivalent of a Josephson Junction'').  

Before moving on, it is worth commenting on the relative scales of the two energies for a Josephson junction, i.e., $E_C$ and $E_J$. Since Josephson junctions are physically very small, the effective capacitance of the junction is often small for qubits (in the fF to low pF  range), while the effective inductance will typically be in the nH to low $\mu$H range \cite{vool2017introduction}. However, through careful engineering of the circuitry around a Josephson junction, great control over both the effective $E_C$ and $E_J$ of the qubit is possible. This can be done using additional Josephson junctions, or by utilizing linear inductors and capacitors (which can be either lumped or distributed elements). Using these tools, the characteristic ratio $E_J/E_C$ can span ranges all the way from less than 0.1 up to $10^6$ depending on a device design \cite{gu2017microwave}. This flexibility has been used to design many different qubits made from various combinations of Josephson junctions and auxiliary circuit elements, leading to a vast trade space for the optimization of qubits for circuit QED systems \cite{gu2017microwave}.

\begin{emphasize}
\begin{center}
	\textbf{Mechanical Equivalent of a Josephson Junction}
\end{center}

\vspace{0.1cm}
{\centering
	\includegraphics[width=0.95\textwidth]{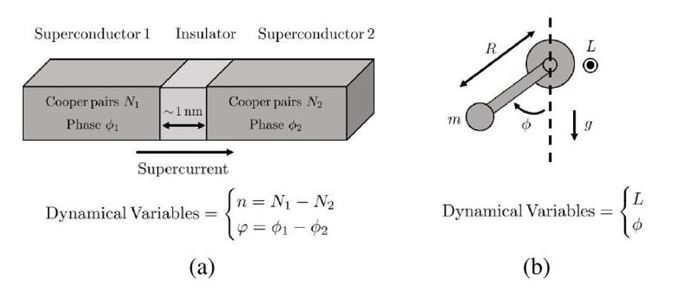}
	\captionof{figure}{Illustrations of (a) a Josephson junction and (b) a simple mechanical pendulum.
		\label{fig:jj-and-pendulum}}
	\par}
\vspace{0.1cm}

There exists a simple mechanical analogue to the Josephson junction that helps build intuition into its dynamics \cite{goldstein1980classical}. In particular, it can be shown that the Hamiltonian for a pendulum of mass $m$ attached to its center of rotation by a rigid, massless rod of length $R$ is mathematically equivalent to (\ref{eq:classical-JJ-Hamiltonian}). The comparison of these two systems is detailed in Fig. \ref{fig:jj-and-pendulum}. The Hamiltonian for the pendulum system is
\begin{align}
H = \frac{1}{2mR^2}L^2 - mgR \cos{\phi},
\label{eq:pendulum-hamiltonian}
\end{align}
where $L$ is the angular momentum of the pendulum, $g$ is the gravitational acceleration, and $\phi$ is the angular position of the pendulum's mass. The resting position of the mass (which is aligned with the gravitational field) is given by $\phi = 0$. Comparing (\ref{eq:classical-JJ-Hamiltonian}) and (\ref{eq:pendulum-hamiltonian}), it is seen that the Josephson junction variables $\varphi$ and $n$ can be identified with the angular position of the pendulum's mass and its angular momentum, respectively.

The Hamiltonians of these systems can be used to derive equations of motion for the Josephson phase and the angular position of the pendulum's mass. For the Josephson junction, we have
\begin{align}
\frac{d^2 \varphi}{dt^2} + 8 E_C E_J \sin\varphi = 0,
\end{align}
while for the mechanical system we have
\begin{align}
\frac{d^2\phi}{dt^2} + \frac{g}{R}\sin\phi = 0.
\label{eq:pendulum-eom}
\end{align}
Recalling that $E_C$ is inversely proportional to the junction capacitance, we can see that the junction capacitance impacts the dynamics of the Josephson phase in a similar manner to the radius of the mechanical pendulum. 

The transmon corresponds to a qubit designed to force the dynamics of the Josephson junction to keep $\varphi$ small. In this regime, the $\sin\varphi$ term can be expanded in its Taylor series. The transmon is operated in a regime where the first two non-zero terms of the Taylor series provide a good approximation to the overall system's dynamics (this is often referred to as a weakly anharmonic oscillator \cite{goldstein1980classical}). As will be discussed in detail in the main text, this regime is reached by using clever engineering to lower $E_C$ by increasing the effective capacitance of the Josephson junction. From the view of the mechanical analogue, this change is similar to making $R$ larger while keeping $g$ fixed.
\end{emphasize}

\subsection{Magnetic Flux-Tunable Josephson Junction}
\label{subsec:jj-operating-regimes}
When Josephson junctions are used to make qubits, it can be advantageous to be able to tune the operating characteristics of the overall system. One common modification used with transmon qubits is to use two similar Josephson junctions in parallel as opposed to using a single junction \cite{koch2007charge,kockum2019quantum}. The two junctions then form a superconducting loop, which is often referred to as a superconducting quantum interference device (SQUID). The SQUID arrangement allows the effective Josephson energy of the overall circuit to be tuned through the application of an external magnetic flux \cite{koch2007charge}. By modifying the Josephson energy, the effective inductance of the SQUID loop can be tuned dynamically. This makes unique qubit operations possible that are not achievable with fixed qubits, such as natural atoms or ions \cite{lang2013correlations}. 

To see why this occurs, we must revisit the effective inductance Hamiltonian ${H}_L$ of the Josephson junction. With the second junction, this part of the Hamiltonian becomes
\begin{align}
{H}_L = -E_J \cos{\varphi}_1 - E_J\cos{\varphi}_2,
\label{eq:split-inductance}
\end{align} 
where ${\varphi}_i$ is the Josephson phase of the $i$th junction and we have assumed that the Josephson energy for both junctions are identical. In reality, the Josephson energies are often made asymmetric on purpose to limit the tuning range of the SQUID. However, junction asymmetries do not change the overall conceptual result, so they are ignored here for simplicity \cite{koch2007charge}.

In addition to (\ref{eq:split-inductance}), the physics of superconductors places another constraint on the relationship between ${\varphi}_1$ and ${\varphi}_2$. In particular, because the phase of the collective wavefunction describing the superconducting condensate is single-valued, it is necessary that the total phase difference around a superconducting loop be an integer multiple of $2\pi$ \cite{langford2013circuit}. However, it can also be shown that an applied magnetic flux intersecting the superconducting loop will affect the total phase change of the condensate wavefunction around the loop (this can be viewed in a manner similar to Faraday's law of induction or the Aharonov-Bohm effect). Overall, the result of this is that the Josephson phase difference around the SQUID is
\begin{align}
{\varphi}_1 - {\varphi}_2 = 2\pi \ell + 2\pi \Phi/\Phi_0,
\label{eq:flux-quantization}
\end{align}
where $\ell$ is an integer, $\Phi$ is the total magnetic flux intersecting the loop formed by the SQUID, and $\Phi_0 = h/2e$ is the superconducting flux quantum \cite{koch2007charge}.

Now, we can utilize standard trigonometric identities to rewrite (\ref{eq:split-inductance}) as
\begin{align}
{H}_L = -2 E_J \cos\bigg(  \frac{{\varphi}_1 -{\varphi}_2}{2} \bigg) \cos\bigg(  \frac{{\varphi}_1 + {\varphi}_2}{2} \bigg).
\end{align}
Defining a new Josephson phase as ${\varphi} = ({\varphi}_1 + {\varphi}_2)/2$ and using (\ref{eq:flux-quantization}), we can simplify the above to be
\begin{align}
{H}_L = - E_{J\Phi} \cos{\varphi},
\label{eq:split-inductance-simplified}
\end{align}
where $E_{J\Phi} = 2 E_J \cos(\pi \Phi / \Phi_0)$. 

From this, we see that the effective inductance of the SQUID has the same form as that of a single Josephson junction. The main change is that the effective Josephson energy $E_{J\Phi}$ is now tunable due to an applied magnetic flux. Although this added control mechanism has many advantages, it also introduces a new way for a qubit with a SQUID to decohere due to interactions with a ``noisy environment''. As a result, the loop size of the SQUID is often extremely small (typically $O(10\,\mu\mathrm{m})$ on each side of the loop) and the qubits are kept inside a shielded enclosure to minimize the sensitivity to environmental flux noise \cite{zhou2020tunable}.

\section{Development of the Transmon Qubit}
\label{sec:transmon-dev}
\begin{figure}[t]
	\centering
	\begin{subfigure}[t]{0.3\linewidth}
		\includegraphics[width=\textwidth]{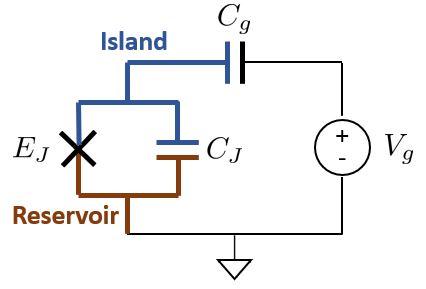}
		\caption{}
		\label{subfig:basic_cpb}
	\end{subfigure}
	\begin{subfigure}[t]{0.27\linewidth}
		\includegraphics[width=\textwidth]{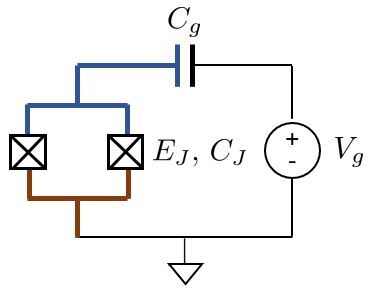}
		\caption{}
		\label{subfig:split_cpb}
	\end{subfigure}
	\begin{subfigure}[t]{0.38\linewidth}
		\includegraphics[width=\textwidth]{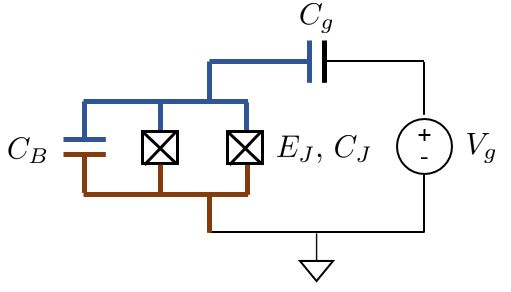}
		\caption{}
		\label{subfig:transmon}
	\end{subfigure}
	\caption{Circuit schematics for the evolution of a CPB to a transmon. (a) CPB, (b) split CPB, and (c) transmon. A pure Josephson junction tunneling element is represented schematically as an ``X''. Often, to simplify the schematics the small junction capacitance $C_J$ is absorbed into the Josephson tunneling element symbol and is represented as a box with an ``X'' through it, as seen in (b) and (c).}
	\label{fig:cpb_schematics}
\end{figure}

One of the earliest qubits used to observe macroscopic quantum behavior in circuit QED systems was the Cooper pair box (CPB) \cite{nakamura1999coherent}, which can be viewed as a predecessor to the transmon. Hence, to better understand the transmon, it is useful to consider it in the context of how it evolved from the simpler CPB qubit. This evolution is shown in terms of circuit schematics in Fig. \ref{fig:cpb_schematics}. We will review each of these different qubits in the coming sections.

\subsection{Introduction to the Cooper Pair Box}
\label{subsec:cpb}
The traditional CPB is formed by a Josephson junction that connects a superconducting ``island'' and ``reservoir'' \cite{kockum2019quantum}. For the CPB, the island is not directly connected to other circuitry, while the reservoir can be in contact with external circuit components (if desired). Since the superconducting island is isolated from other circuitry, the CPB is very sensitive to the number of Cooper pairs that have tunneled through the Josephson junction (down to a single Cooper pair). Due to this sensitivity, the CPB is also called a charge qubit \cite{kockum2019quantum}.

Typically, it is advantageous to be able to control the operating point of the CPB system. For instance, this can help with initializing the state of the CPB before conducting an experiment. Generally, the operating point of the CPB is set by a voltage source capacitively coupled to the superconducting island. This voltage can be tuned to induce a certain number of ``background'' Cooper pairs that have tunneled onto the superconducting island \cite{kockum2019quantum}. The basic circuit diagram of this qubit is shown in Fig. \ref{subfig:basic_cpb}. Although the circuit is changed somewhat compared to a simple Josephson junction, the resulting Hamiltonian still has a very similar form to (\ref{eq:classical-JJ-Hamiltonian}). In particular, the CPB Hamiltonian is 
\begin{align}
H = 4E_C (n-n_g)^2 - E_J \cos\varphi,
\label{eq:classical-CPB}
\end{align}
where $n_g$ is the offset charge induced by the voltage source and $E_C$ must be modified to account for the effect of $C_g$ \cite{girvin2011circuit}.

At this point, it is useful to consider the quantum description of the CPB (for an introduction to quantum mechanical principles, see \cite{chew2021qme-made-simple}). Since we have a Hamiltonian description of the system already expressed in terms of conjugate variables in (\ref{eq:classical-CPB}), finding a quantum description of the system is quite simple \cite{chew2021qme-made-simple}. In particular, the two conjugate variables are elevated to be non-commuting quantum operators denoted as $\hat{n}$ and $\hat{\varphi}$, which can be viewed as being synonymous with infinite-dimensional matrix operators. These operators must always satisfy the commutator $[\hat{\varphi},\hat{n}] \! =  \hat{\varphi}\hat{n} - \hat{n}\hat{\varphi} = \! i$ \cite{vool2017introduction}. Combined with a complex-valued quantum state function, denoted as $|\psi\rangle$ in Dirac notation, these quantum operators take on a statistical interpretation and share an uncertainty principle relationship \cite{chew2021qme-made-simple}. In particular, the uncertainty principle for $\hat{n}$ and $\hat{\varphi}$ gives that the standard deviations of the two operators must follow $\sigma_n \sigma_\varphi \geq 1/2$. As a result, measurements of observables associated with these quantum operators (e.g., the number of Cooper pairs that have tunneled through the junction) cannot be measured simultaneously with arbitrary precision \cite{chew2021qme-made-simple}.

Now, in transitioning to a quantum description the classical Hamiltonian for the CPB given in (\ref{eq:classical-CPB}) becomes a Hamiltonian operator given by
\begin{align}
\hat{H} = 4 E_C (\hat{n}-n_g)^2 - E_J \cos \hat{\varphi},
\label{eq:quantum-CPB}
\end{align}
where it is noted that $n_g$ remains a classical variable that describes the offset charge induced by an applied DC voltage. This Hamiltonian can be used in the quantum state equation (a generalization of the Schr\"{o}dinger equation) to determine the time evolution of a quantum state \cite{chew2021qme-made-simple}. In particular, the quantum state equation is
\begin{align}
\hat{H}|\psi\rangle = i\hbar \partial_t |\psi \rangle.
\label{eq:quantum-state-equation}
\end{align}
In the process of solving (\ref{eq:quantum-state-equation}), it is often useful to identify the stationary states of the Hamiltonian by converting (\ref{eq:quantum-state-equation}) into a time-independent eigenvalue problem. This is
\begin{align}
\hat{H}|\Psi\rangle = E |\Psi \rangle,
\label{eq:quantum-state-stationary}
\end{align}
where $E$ is the energy associated with stationary state (or eigenvector) $|\Psi\rangle$.

For a typical CPB, $E_J/E_C < 1$. As a result, it becomes useful to write the qubit Hamiltonian given in (\ref{eq:quantum-CPB}) in terms of \textit{charge states}. These are eigenstates of $\hat{n}$, and are often denoted as $|N\rangle$ (in this notation, $|N\rangle$ can be viewed like a column vector, with $\langle N |$ being its conjugate transpose). The eigenvalue associated with this eigenstate is $N$, which is always a discrete number that counts how many Cooper pairs have tunneled through the Josephson junction relative to the equilibrium state. In terms of charge states, the effective capacitance term of the CPB Hamiltonian becomes
\begin{align}
4E_C(\hat{n}-n_g)^2 = 4E_C \sum_N (N-n_g)^2 |N \rangle \langle N|.
\end{align}
The combination $|N\rangle\langle N |$ can be viewed as an outer product between the states $|N\rangle$ and $|N\rangle$. Hence, the operator $|N\rangle\langle N |$ can be viewed like a matrix in a more familiar linear algebra notation. Since the charge states are all orthogonal, the sum of operators $|N\rangle\langle N |$ can be viewed as ``diagonalizing'' this term of the Hamiltonian. Now, in terms of charge states the effective nonlinear inductance term of the CPB Hamiltonian becomes \cite{vool2017introduction}
\begin{align}
-E_J \cos\hat{\varphi} = \frac{E_J}{2} \sum_N \big( |N\rangle\langle N+1 | + |N+1\rangle\langle N|    \big).
\end{align}
Due to the orthogonality of the charge states, the operator $|N\rangle\langle N+1 |$ represents the coherent transfer of a single Cooper pair from the island to the reservoir. Correspondingly, the operator $|N+1 \rangle\langle N |$ has a similar interpretation, but with the Cooper pair being transferred from the reservoir to the island. Hence, this representation clearly shows the tunneling physics involved in the effective inductance of the CPB.

\subsection{Cooper Pair Box as a Qubit}
To determine how a CPB can be used as a qubit, it is useful to consider its operating characteristics as a function of the offset charge $n_g$ (for a discussion on the needed properties of a qubit see ``Essential Properties of a Qubit''). As we will see shortly, it is advantageous to set $n_g$ to either integer or half-integer values for typical qubit operations \cite{kockum2019quantum}. For this discussion, the primary benefits of operating at these $n_g$ points can be best understood by considering the energy level diagram of a CPB as a function of $n_g$. This diagram plots the energy eigenvalues $E_i$ of (\ref{eq:quantum-state-stationary}) for different eigenvectors (or stationary states) $|\Psi_i \rangle$. The operating frequency of the CPB qubit (i.e., what frequencies of microwave radiation can be absorbed or emitted) is based on the energy difference between two adjacent energy levels. 

\begin{figure}[t]
	\centering
	\includegraphics[width=0.7\linewidth]{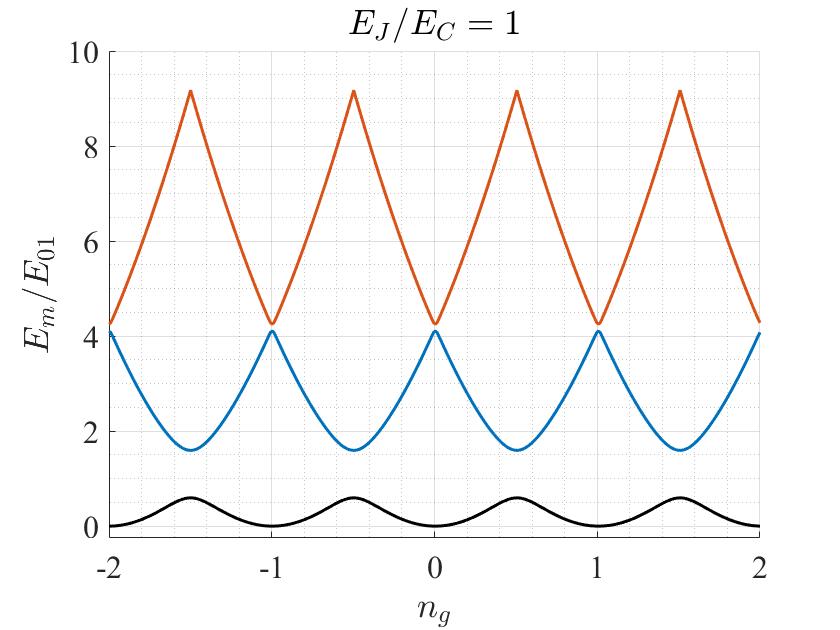}
	\caption{First three energy levels of the qubit Hamiltonian given in (\ref{eq:quantum-CPB}) for $E_J/E_C = 1$. Energy levels are normalized by the transition energy between the first two states evaluated at half-integer values of $n_g$.}
	\label{fig:energy_levels}
\end{figure}

The energy level diagram is shown for the first three energy levels of the CPB in Fig. \ref{fig:energy_levels}, where it is seen that the energy levels strongly depend on $n_g$. Beginning with $n_g$ having an integer value, it is seen that there is a very large separation between the ground and excited states. This large spacing makes it highly unlikely that the CPB can be spontaneously excited, making this operating point useful for quickly initializing the CPB to its ground state.

Although this operating point can establish a good ground state, it is not useful for qubit operations because the first two excited states have similar energies. This makes it difficult to prevent the system from unintentionally transitioning into and out of the second excited state once the first excited state has been populated. Hence, a different operating point is needed to perform qubit operations.

Considering the energy level diagram of Fig. \ref{fig:energy_levels} again, it is seen that half-integer values of $n_g$ are ideal for performing qubit operations. At these values, there is large separation between the energy levels of the first two excited states. Further, it is seen that the energy levels exhibit a high degree of anharmonicity here (i.e., they are not evenly spaced like a harmonic oscillator or linear resonator). As a result of these two properties, the transition between the ground and first excited state can be selectively driven with fast microwave pulses (on the order of 10 ns \cite{koch2007charge,motzoi2009simple}) without requiring significant filtering of the pulse. This allows for qubit operations to be performed very quickly, which is one factor that has made these devices attractive for building a quantum computer. 

The flexibility of operating a CPB can be increased further by replacing the single Josephson junction with a SQUID. This configuration is known as a split CPB, and is shown schematically in Fig. \ref{subfig:split_cpb}. By applying a magnetic flux, the $E_J$ of the split CPB can be tuned dynamically. Since the energy difference between different states seen in Fig. \ref{fig:energy_levels} depends on $E_J$, the SQUID can be used to tune the operating frequency of the split CPB \textit{in situ}.

A further benefit to operating at half-integer values of $n_g$ is that the two lowest energy levels are locally flat near this operating point (i.e., the slope is approximately $0$). As a result, the qubit transition frequency is less sensitive to noise in the $n_g$ variable (often referred to as charge fluctuations or charge noise). The improved coherence of the CPB at this operating point has led to it being known as the ``sweet spot'' for CPBs \cite{koch2007charge}. Using these strategies, CPBs with coherence times on the nanosecond to microsecond scale were able to be achieved \cite{devoret2013superconducting}. Although this was sufficient to replicate various quantum phenomena observed in quantum optical systems \cite{gu2017microwave}, the CPB was still too sensitive to charge noise to form a part of a scalable quantum information processing system. The transmon was introduced to address this issue.

\begin{emphasize}
	\begin{center}
		\textbf{Essential Properties of a Qubit}
	\end{center}	
	\vspace{0.1cm}
	For a physical system to be used as a qubit it must have two distinct quantum states that can be ``isolated'' from other states in the system. Further, it is necessary to be able to selectively drive transitions between the two states that form the qubit without accidentally exciting other states of the physical system. For the qubits discussed in this review, a qubit is formed by the two lowest energy levels of the system. The difference in energy levels dictates what frequencies of microwave radiation can be absorbed or emitted from the qubit. As a result, it is necessary for the energy levels to be unevenly spaced for the physical system to act as a qubit.
	
	Oftentimes, this is expressed as needing the qubit to be nonlinear. The nonlinearity is in reference to the ``restoring force'' due to the potential energy term of the Hamiltonian. Physical systems with a quadratic potential lead to linear restoring forces, and are characterized by having evenly spaced energy levels (these are referred to as quantum harmonic oscillators). Linear transmission lines and LC resonators have these characteristics, and as a result cannot be used as qubits. In contrast to this, the CPB and transmon have a cosine potential energy term that leads to a nonlinear restoring force and unevenly spaced energy levels (see Fig. \ref{fig:potentials-illustrations}). Hence, they are candidates to be used as qubits.
	
	It is also necessary to be able to reliably initialize the state of the qubit to a known value, often the qubit's ground state. Generally, this requires the qubit to be isolated from any environments that could provide enough energy to the qubit to spontaneously raise it to its excited state. One prevalent source of this is thermal energy, and is one of the driving reasons for why circuit QED systems must be operated at such low temperatures. 
	\vspace{0.05cm}
	
	{\centering
	\includegraphics[width=\textwidth]{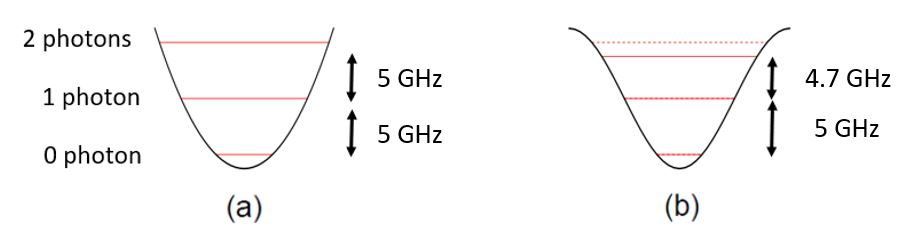}
	\captionof{figure}{Potential energies and corresponding energy level spacing for (a) a quadratic potential (e.g, that of a linear LC resonator) and (b) a cosine potential (e.g., that of a nonlinear resonator like the CPB or transmon). In (b), the corresponding levels of a harmonic oscillator are included as dashed lines to highlight the anharmonicity of the cosine potential.
	\label{fig:potentials-illustrations}}
	\par}
\end{emphasize}

\subsection{Transmon}
The transmon qubit is best viewed as a CPB shunted by a capacitor that is large relative to the stray capacitance of the Josephson junction \cite{koch2007charge}. This is shown schematically in Fig. \ref{subfig:transmon}, where $C_B$ is the large shunting capacitance. Recalling that $E_C$ is inversely proportional to capacitance, this results in the characteristic ratio $E_J/E_C$ transitioning from $E_J/E_C < 1$ for a traditional CPB to $E_J/E_C \gg 1$ for a transmon.

\begin{figure}[t]
	\centering
	\includegraphics[width=\linewidth]{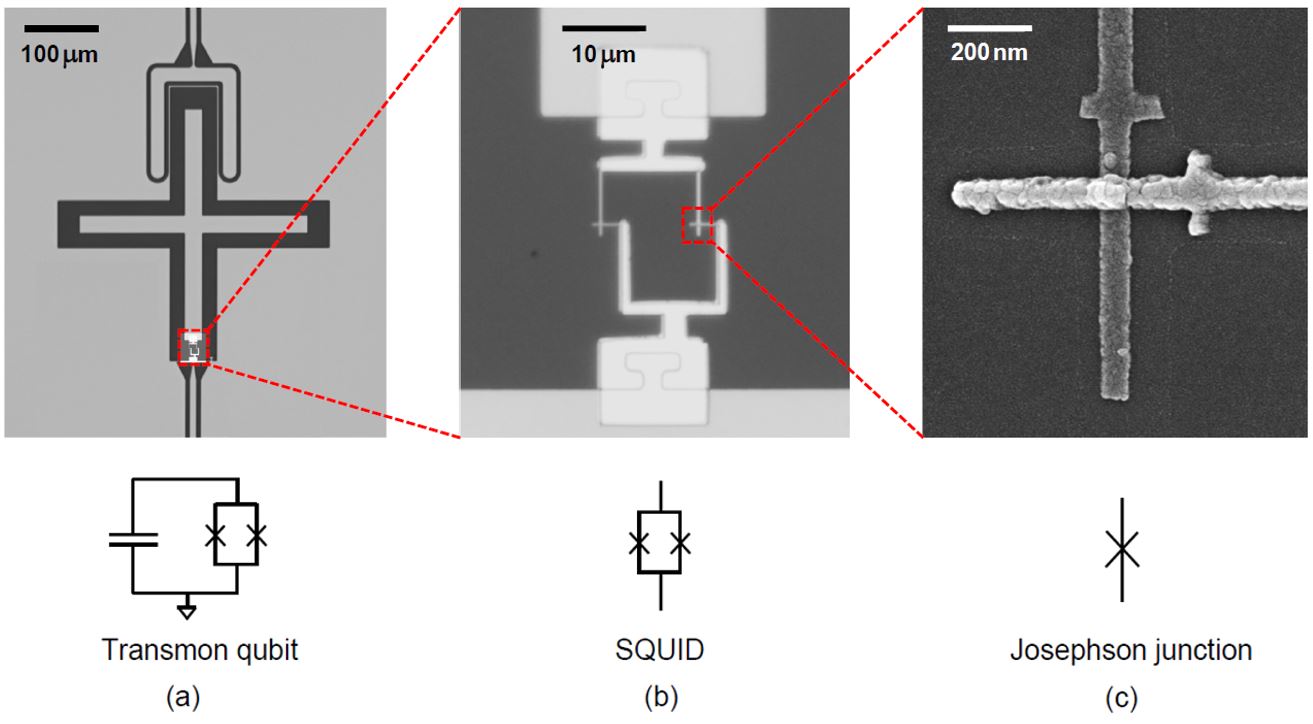}
	\caption{Images of a frequency tunable transmon qubit. (a) The entire transmon consisting of a large capacitance (the ``+'' shape) in parallel with a SQUID to ground. A zoom in of the SQUID and a single Josephson junction are in (b) and (c), respectively.}
	\label{fig:physical_transmon}
\end{figure}

Originally, the large shunting capacitance was made using interdigital capacitors \cite{koch2007charge}. Now, the shunting capacitance is more commonly implemented as a large ``cross'' shape (this is shown in Fig. \ref{fig:physical_transmon}, and is sometimes also referred to as an ``Xmon qubit'') \cite{barends2014superconducting}. This newer shape is typically favored as it provides better interconnectivity between different parts of a design. Although the physical implementation of the transmon qubit is different from more traditional CPB qubits, a Hamiltonian with the same form as (\ref{eq:quantum-CPB}) can still be used to describe its behavior \cite{koch2007charge}. Similarly, a SQUID can be used to connect the different superconductors that make up the transmon (see Fig. \ref{fig:physical_transmon}), allowing the operating frequency of the transmon to be tuned via an applied magnetic flux.

\begin{figure}[t]
	\centering
	\begin{subfigure}[t]{0.49\linewidth}
		\includegraphics[width=\textwidth]{Ej_Ec_1}
		\caption{}
		\label{subfig:energy1}
	\end{subfigure}
	\begin{subfigure}[t]{0.49\linewidth}
		\includegraphics[width=\textwidth]{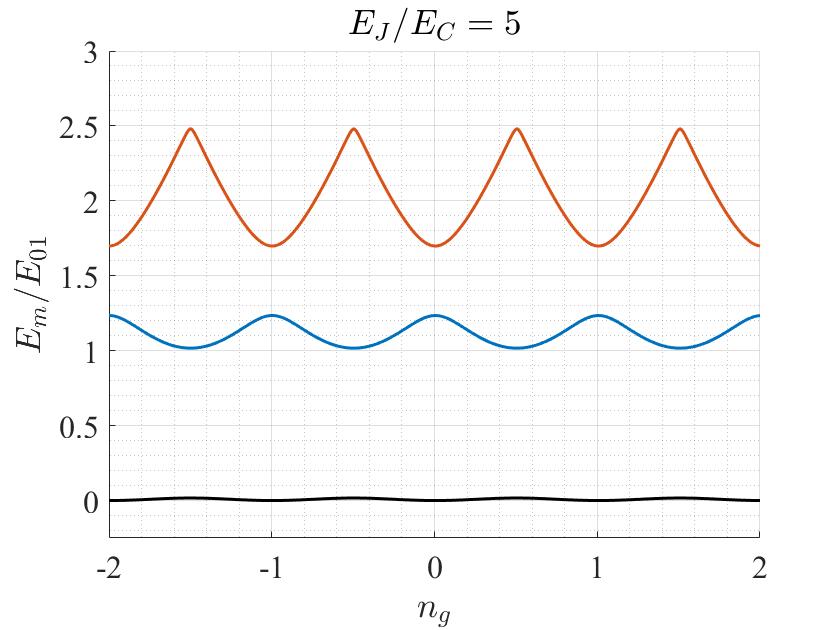}
		\caption{}
		\label{subfig:energy2}
	\end{subfigure}
	\begin{subfigure}[t]{0.49\linewidth}
		\includegraphics[width=\textwidth]{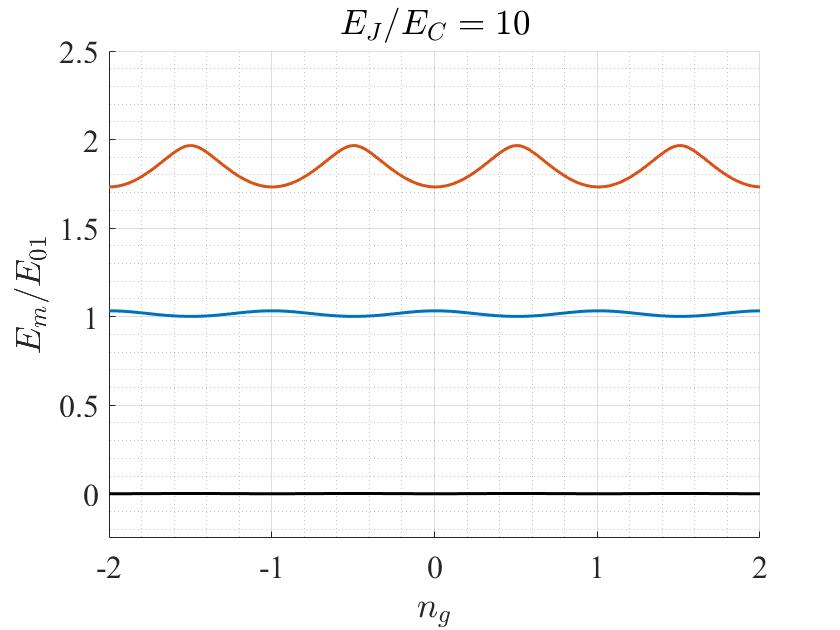}
		\caption{}
		\label{subfig:energy3}
	\end{subfigure}
	\begin{subfigure}[t]{0.49\linewidth}
		\includegraphics[width=\textwidth]{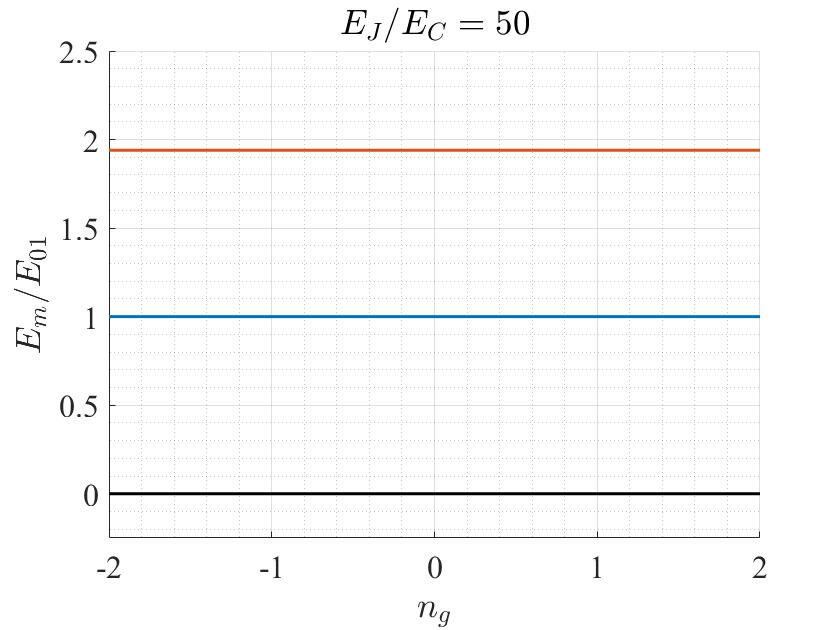}
		\caption{}
		\label{subfig:energy4}
	\end{subfigure}
	\caption{First three energy levels of the qubit Hamiltonian given in (\ref{eq:quantum-CPB}) for values of $E_J/E_C$ ranging from the CPB regime in (a) to the transmon regime in (d). Energy levels are normalized by the transition energy between the first two states evaluated at half-integer values of $n_g$.}
	\label{fig:energy_levels2}
\end{figure}

It is instructive to see how the first few energy levels of (\ref{eq:quantum-CPB}) change as a function of $E_J/E_C$ to understand the impact of adding the large shunting capacitance around the Josephson junction. This is shown for four different values of $E_J/E_C$ in Fig. \ref{fig:energy_levels2}. These demonstrate the transition from a traditional CPB in Fig. \ref{subfig:energy1} to a transmon in Fig. \ref{subfig:energy4}. From Fig. \ref{fig:energy_levels2}, it is seen that as $E_J/E_C$ is increased, the energy levels flatten out as a function of $n_g$ while simultaneously becoming more harmonic (closer to equally spaced like a quantum harmonic oscillator). Fortunately, the sensitivity to charge noise reduces at an exponential rate, while the anharmonicity reduces following a weak power law \cite{koch2007charge}. This leads to a qubit that is insensitive to charge noise for practical purposes, but is still able to provide sufficient anharmonicity to effectively perform qubit operations with short-duration microwave pulses. Typically, the main operating frequencies of transmons range from a few GHz to 10 GHz, with anharmonicities of \textapprox100 to 300 MHz.

From the view of a mechanical equivalent, the transmon case is analogous to the rigid pendulum with dynamics dominated by a gravitational field (see ``Mechanical Equivalent of a Josephson Junction'') \cite{koch2007charge}. This forces the angular values of the pendulum to small values. For this operating point, the dynamics correspond to a weakly anharmonic oscillator, which leads to the energy level separation seen in Fig. \ref{subfig:energy4}. The flatness of the energy levels in Fig. \ref{subfig:energy4} can be understood from the transmon Hamiltonian (\ref{eq:quantum-CPB}), where it is seen that $n_g$ is multiplied by $E_C$. Since $E_C$ is made small for a transmon, $n_g$ can only have a small role in the dynamics. Due to this, the voltage source seen in Fig. \ref{subfig:transmon} is no longer a useful control mechanism, and so is omitted for practical designs.

As a result of the transmon's resilience to charge noise (and other manufacturing and experimental advances), the coherence times of modern transmons are able to achieve values in the hundreds of microseconds range, representing an orders of magnitude improvement over the best CPB qubits \cite{devoret2013superconducting,place2021new}. However, other sources of decoherence still persist, making the identification and reduction of decoherence still an important research topic \cite{krantz2019quantum}.
\section{Interfacing Transmon Qubits with Other Circuitry}
\label{sec:transmon-intefacing}
A completely isolated qubit cannot be controlled or measured, and so is of little use. Further, to implement many kinds of quantum information processing operations, multiple qubits that are spatially separated need to interact with each other to generate entanglement. Circuit QED systems address this by coupling qubits to transmission line structures. Often, these transmission lines will be resonators that can be used as a quantum ``bus'' to connect different qubits to each other or as part of a measurement chain to read out the quantum states of the qubits (e.g., see Fig. \ref{fig:full-device}) \cite{blais2004cavity,blais2007quantum,gu2017microwave}. 

\begin{figure}[t]
	\centering
	\begin{subfigure}[t]{0.9\linewidth}
		\includegraphics[width=\textwidth]{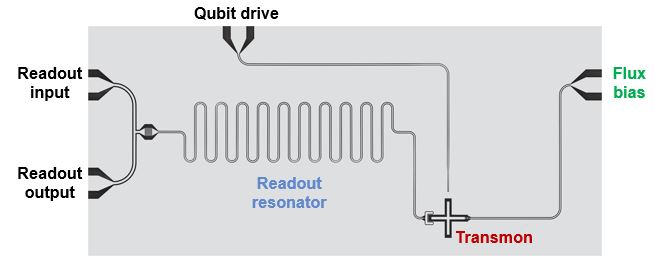}
		\caption{}
		\label{subfig:transmon_and_resonator}
	\end{subfigure}
	\begin{subfigure}[t]{0.63\linewidth}
		\includegraphics[width=\textwidth]{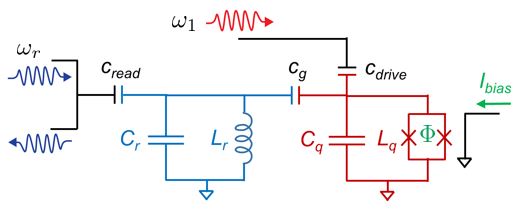}
		\caption{}
		\label{subfig:transmon_and_resonator_schematic}
	\end{subfigure}	
	\begin{subfigure}[t]{0.35\linewidth}
		\includegraphics[width=\textwidth]{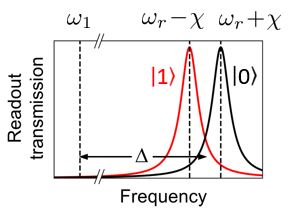}
		\caption{}
		\label{subfig:readout-transmission}
	\end{subfigure}
	\caption{(a) Illustration of a typical single transmon device. Microwave pulses are applied to the qubit drive line to modify the state of the transmon. The flux bias line is used to dynamically tune the operating frequency of the transmon. The state of the transmon (e.g., if it is in its ground or excited state) can be monitored via microwave transmission measurements made between the two readout ports. (b) Equivalent lumped element circuit model of the device in (a). (c) Transmon state-dependent transmission spectra between the readout ports.}
	\label{fig:transmon_and_resonator}
\end{figure}

An example of the typical on-chip features needed to operate a single transmon and the corresponding approximate circuit model are shown in Fig. \ref{fig:transmon_and_resonator}. As is common in the circuit QED community, a single mode of the transmission line resonator is modeled in a lumped element approximation as an LC tank circuit in Fig. \ref{subfig:transmon_and_resonator_schematic}. \cite{koch2007charge}. More modes of the resonator often must be considered to provide more accurate predictions of experimental results, and can be accounted for by including additional LC tank circuits \cite{houck2008controlling,nigg2012black}. Regardless, the lumped element approximation can make it difficult to optimize the design of these systems. Hence, developing full-wave numerical modeling methods that can rigorously analyze these systems is an area of future research interest \cite{roth2021circuit}. 

The main features to operate the transmon shown in Fig. \ref{fig:transmon_and_resonator} are the flux bias line, the qubit drive line, and the state readout features. The flux bias line is used to apply a magnetic flux to the SQUID loop of the transmon to change its operating frequency. This can help correct for manufacturing variability and can also be a key capability in executing certain quantum information processing algorithms \cite{barends2014superconducting}.
	
The qubit drive line is used to modify the state of the qubit by applying a microwave drive pulse with a center frequency that matches the transmon's operating frequency. If the transmon starts in its ground state, an appropriately designed microwave drive pulse can be used to raise the transmon to its excited state or to place it in some superposition of the transmon's ground and excited states \cite{blais2007quantum}. Further microwave drive pulses can be applied along the qubit drive line to modify the transmon's state as needed throughout the course of executing a quantum algorithm.

Finally, it is necessary to be able to determine the state of the transmon after executing a quantum algorithm. The most common way to do this is known as \textit{dispersive readout}, with a simple setup to achieve this shown in Fig. \ref{fig:transmon_and_resonator} \cite{blais2007quantum}. This approach leverages that the state of the transmon (e.g., if it is in its ground or excited state) modifies the resonant frequencies of the readout resonator, as shown in Fig. \ref{subfig:readout-transmission}. These changes can be monitored via microwave spectroscopy in the form of transmission measurements made between the readout ports.

All of these effects (and many more) can be studied for circuit QED systems using a generalization of the famous Jaynes-Cummings Hamiltonian, which is one of the most widely studied models in quantum optics and cavity QED \cite{gerry2005introductory}. This is also the starting point for many applied circuit QED studies. To help transition between this work and the modern literature, we now provide a brief introduction to how this Hamiltonian can be used for circuit QED systems. 

\subsection{Generalized Jaynes-Cummings Hamiltonian}
To simplify the development, we will only focus on describing the case of a single transmon qubit capacitively coupled to a single resonator. More qubits and resonators can be handled as simple extensions of the model discussed here.

A rigorous development of the Hamiltonian description of this circuit is tedious, and so will be omitted for brevity (for details, see \cite[Appx. A]{girvin2011circuit}). Here, we will focus on an intuitive development only. As is often possible for coupled systems, we can write the total Hamiltonian as a sum of uncoupled (or ``free'') Hamiltonians and an interaction Hamiltonian that accounts for the coupling. Considering this, we will have
\begin{align}
\hat{H} = \hat{H}_T + \hat{H}_R + \hat{H}_I,
\label{eq:coupled-system-hamiltonian}
\end{align}
where $\hat{H}_T$ ($\hat{H}_R$) denotes the free transmon (resonator) Hamiltonian and $\hat{H}_I$ is the interaction Hamiltonian. The free transmon Hamiltonian has already been given in (\ref{eq:quantum-CPB}). However, to further simplify the analysis, the transmon Hamiltonian is typically diagonalized in terms of its eigenstates \cite{koch2007charge}. Using these eigenstates, (\ref{eq:quantum-CPB}) can be rewritten as
\begin{align}
\sum_j \hbar\omega_j |j\rangle\langle j |,
\label{eq:free-transmon}
\end{align}
where $\omega_j$ is the eigenvalue associated with eigenstate $|j\rangle$ and the operator $|j\rangle\langle j |$ can be viewed as part of an eigenmode decomposition of the Hermitian ``matrix'' $\hat{H}_T$.

The free resonator Hamiltonian is simple to write since it is the total energy contained in the LC tank circuit elements $C_r$ and $L_r$. From basic circuit theory, this will be
\begin{align}
\hat{H}_r = \frac{1}{2}[L_r \hat{I}_r^2 + C_r \hat{V}_r^2 ],
\label{eq:free-resonator}
\end{align}
where $\hat{I}_r$ and $\hat{V}_r$ are the current and voltage in the LC tank circuit, respectively. Typically, it is advantageous to express the voltage and current in terms of \textit{ladder operators} that diagonalize $\hat{H}_r$ \cite{koch2007charge}. In particular, $\hat{a}^\dagger$ ($\hat{a}$) is known as the creation (annihilation) operator, and when it operates on the quantum state of the resonator it increases (decreases) the photon number by one \cite{gerry2005introductory}. Using the properties of the ladder operators, (\ref{eq:free-resonator}) can be rewritten as 
\begin{align}
\hat{H}_r =	\hbar\omega_r \hat{a}^\dagger\hat{a},
\end{align}
where $\omega_r$ is the resonant frequency of the transmission line resonator and a constant term (known as the zero point energy) has been removed since it does not affect the dynamics.

One way to approach determining $\hat{H}_I$ is to recognize that because it is a capacitive coupling it will be sensible to write the interaction in terms of the voltages and charges. Now, for many circuit QED systems the resonator voltage as seen from the transmon can be approximately viewed as coming from an ideal voltage source. Considering this, the interaction between the resonator and transmon can be given by
\begin{align}
\hat{H}_I = 2e\beta \hat{V}_r \hat{n},
\end{align}
where $\hat{n}$ is the charge operator of the transmon and $\beta = C_g/(C_g+C_q)$ is a capacitive voltage divider that places the correct ratio of the resonator voltage on the transmon \cite{koch2007charge}. In terms of the resonator ladder operators, this becomes
\begin{align}
\hat{H}_I = 2e\beta  V_\mathrm{rms} (\hat{a}+\hat{a}^\dagger) \hat{n},
\label{eq:H-int}
\end{align}
where $V_\mathrm{rms} = \sqrt{\hbar\omega_r/2C_r}$. To further simplify the analysis, $\hat{n}$ can be rewritten in terms of transmon eigenstates and a number of standard approximations can be applied to finally arrive at an interaction Hamiltonian of
\begin{align}
\sum_{j}  g_j \big(|j-1\rangle \langle j | \hat{a}^\dagger + |j\rangle \langle j-1 | \hat{a} \big) ,
\end{align} 
where $g_{j} =  2e\beta V_\mathrm{rms} \langle j-1 | \hat{n} | j \rangle $ \cite{koch2007charge}.

Putting all of these results together, the complete system Hamiltonian of (\ref{eq:coupled-system-hamiltonian}) becomes
\begin{multline}
\hat{H} = \sum_j \hbar\omega_j |j\rangle\langle j | + \hbar\omega_r \hat{a}^\dagger\hat{a} \\ +\sum_{j}  g_j \big(|j-1\rangle \langle j | \hat{a}^\dagger + |j\rangle \langle j-1 | \hat{a} \big).
\label{eq:coupled-transmon5}
\end{multline}
This is a generalized form of the Jaynes-Cummings Hamiltonian, which can be used to study many practical effects related to quantum information processing \cite{blais2004cavity,blais2007quantum,gu2017microwave}. In (\ref{eq:coupled-transmon5}), the terms multiplied by $g_j$ represent the coherent exchange of excitations between the transmon and resonator. This ``swapping'' of excitations is particularly prevalent when $\omega_j$ and $\omega_r$ are nearly equal, and is usually referred to as vacuum Rabi oscillations. 

However, when $\Delta = |\omega_j - \omega_r | \gg g_j$, the transmon and resonator are said to be in the \textit{dispersive regime} of circuit QED \cite{gu2017microwave}. In this regime, the Hamiltonian of (\ref{eq:coupled-transmon5}) can be approximated as
\begin{multline}
\hat{H} \approx \hbar\big( \omega_r - \chi |1\rangle\langle 1| + \chi |0\rangle\langle 0|   \big) \hat{a}^\dagger \hat{a} \\ + \frac{\hbar}{2}\big(\omega_1 + \chi \big)\big(|0 \rangle \langle 0| - |1\rangle \langle 1| \big),
\label{eq:dispersive-JC}
\end{multline}
where only the two lowest transmon states have been used for clarity and $\chi = g_1^2/|\omega_1-\omega_r|$ \cite{gu2017microwave}. One important property of (\ref{eq:dispersive-JC}) is that the resonator frequency depends on the state of the transmon (this is seen in the first set of parentheses in (\ref{eq:dispersive-JC})). In particular, the resonator frequency shifts to $\omega_r + \chi$ or $\omega_r - \chi$ if the transmon is in its ground or excited state, respectively. For this reason, $\chi$ is often referred to as the \textit{dispersive shift}. This shift in the resonant frequency can be measured via microwave spectroscopy of the readout resonator. This technique is known as dispersive readout because there is no energy transfer between the transmon and the resonator, and is one of the most common ways to monitor the state of a transmon \cite{gu2017microwave}.

\section{Conclusions and Outlook}
\label{sec:conclusion}
Circuit QED systems have emerged as one of the most promising candidates for quantum information processing on the scale needed to achieve a quantum advantage in practical applications. Although many different qubits have been used in circuit QED systems, the transmon has become the workhorse qubit in superconducting circuit systems.

In this work, we presented a review of the essential properties of the transmon qubit in a manner that is largely accessible to the classical electromagnetic engineering community who have a limited background in quantum mechanics. These systems require a significant amount of classical engineering to be successful, such as designing the classical control and readout systems for the qubits and developing compact electrical components that can operate at the various cryogenic stages of a dilution refrigerator, to name a few. There is also a need for improved classical simulation tools to assist in the design process of these systems. We firmly believe that members of the classical electromagnetic engineering community can play a vital role in the advancement of this new era of electromagnetic devices, and look forward to seeing this happen.



\ifCLASSOPTIONcaptionsoff
  \newpage
\fi

\bibliographystyle{IEEEtran}
\bibliography{CEM_bib}







\end{document}